\documentclass[%
 reprint,
nofootinbib,
 amsmath,amssymb,
 aps,
 prl,
]{revtex4-1}

\usepackage{float}
\usepackage[colorlinks = true, citecolor = blue, linkcolor = blue ]{hyperref}
\usepackage{graphicx}% Include figure files
\usepackage{dcolumn}% Align table columns on decimal point
\usepackage{bm}% bold math
\newcommand{\jcap}{JCAP}
\newcommand{\APh}{APh}
\newcommand{\mnras}{MNRAS}
\newcommand{\aap}{A\&A}

\def \nism {n_{\rm ISM}}
\def \rhoism {\rho_{\rm ISM}}
\def \vsh {v_{\rm sh}}
\def \rl {r_{\rm L}}
\def \xiinj {\xi_{\rm inj}}
\def \Emax {E_{\rm max}}
\def \jcap {JCAP}
\def \aap {A\&A}
\def \APh {APh}
\def \mnras {MNRAS}

\begin{document}

\preprint{APS/123-QED}

\title{On the Spectrum of Electrons Accelerated in Supernova Remnants}

\author{Rebecca Diesing}
\author{Damiano Caprioli}%
\affiliation{%
Department of Astronomy and Astrophysics, University of Chicago, Chicago, IL 60637, USA}%

\begin{abstract}
Using a semi-analytic model of non-linear diffusive shock acceleration, we model the total spectrum of cosmic ray (CR) electrons accelerated by supernova remnants (SNRs). Because electrons experience synchrotron losses in the amplified magnetic fields characteristic of SNRs, they exhibit substantially steeper spectra than protons. In particular, we find that the difference between the electron and proton spectral index (power law slope) ranges from 0.1 to 0.4. Our findings must be reckoned with theories of Galactic CR transport, which often assume that electrons and protons are injected with the same slope, and may especially have implications for the observed ``positron excess."
\end{abstract}

\maketitle

%\section{\label{sec:level1}Introduction}
{\it Introduction.---\label{sec:level1}}
Developing a complete paradigm for the origin of Galactic cosmic rays (CRs) with energies up to $\sim 10^8$ GeV requires a detailed understanding of their acceleration and propagation. The best source candidates for such acceleration are supernova remnants (SNRs), which provide sufficient energetics and an efficient acceleration mechanism \citep{hillas05,caprioli+10a}. 
Namely, particles are scattered by magnetic field perturbations, resulting in diffusion across the SNR forward shock and an energy gain with each crossing \citep[]{fermi54, krymskii77, axford+77, bell78a, blandford+78}. This mechanism, known as \emph{diffusive shock acceleration} (DSA), predicts power law energy distributions of CRs, $\propto E^{-q}$, where $q$ depends only on the shock dynamics and $q\to 2$ for strong shocks.

Once accelerated, CR protons and electrons diffuse through the Galaxy such that their spectrum is modified by escape from the Galaxy and, in the case of electrons, energy losses due to synchrotron and inverse-Compton scattering. 
Moreover, CR protons interact with protons in the interstellar medium (ISM) to produce secondary particles, most notably positrons and antiprotons. Thus, in this standard picture, one would expect the positron and antiproton spectra to follow that of their parent protons, modulo effects due to their subsequent escape and energy loss.

To put this analysis into more quantitative terms, consider CR protons and electrons injected by SNRs with spectra $\propto E^{-q_{\rm p}}$ and $\propto E^{-q_{\rm e}}$, respectively. In the standard paradigm for CR transport, the Galactic residence time of these particles scales as $E^{-\delta}$, with $\delta\sim 0.2-0.4$ \citep[e.g.,][]{ams16b,lipari18}. Thus, we would expect the observed proton spectrum to go as $E^{-(q_{\rm p}+\delta)}$. Leptons also experience energy losses due interactions with the Galactic magnetic field (synchrotron) and radiation fields (inverse Compton). We therefore expect $N_{\rm e} \propto E^{-(q_{\rm e}+\beta)}$, where $\beta\gtrsim \delta$ reflects the effective spectral steepening due to a combination of escape and energy loss. Assuming positrons and antiprotons are secondaries produced in interactions between CR and ISM protons, the positron spectrum should scale as $ E^{-(q_{\rm p}+\delta+\beta)}$, and the antiproton spectrum  as $E^{-(q_{\rm p}+2\delta)}$.

A notable observation that appears to be in conflict with this paradigm is the ``positron excess" observed by PAMELA \citep{pamela13} and AMS-02 \citep{ams14}. Both collaborations report a positron fraction, $\chi \equiv \Phi_{e^+}/(\Phi_{e^-}+\Phi_{e^+})$ where $\Phi_{e^+}$ and $\Phi_{e^-}$ are the positron and electron fluxes, that rises with energy. In the picture described above, $\chi \simeq \Phi_{e^+}/\Phi_{e^-} \propto E^{-(q_{\rm p}-q_{\rm e}+\delta)}$. Thus, under the standard assumption that $q_{\rm p} = q_{\rm e}$, $\chi \propto E^{-\delta}$ and should decrease with energy  \citep[see, e.g.,][for a thorough review]{amato+17}.

This discrepancy may be at least partially resolved if electrons are injected into the Galaxy with a steeper spectrum than protons (i.e., $q_{\rm e} > q_{\rm p}$). Such steepening is physically motivated, as electrons experience synchrotron losses during the acceleration process. Although the lifetime of a typical SNR is much shorter than the CR galactic residence time, CR acceleration leads to magnetic field amplification \citep{skilling75a, bell78a,bell04, amato+09}, producing magnetic fields hundreds of times stronger than that of the Galaxy \cite[e.g.,][]{volk+05, caprioli+09a}. The result is that the synchrotron loss time in SNRs is generally shorter than the DSA timescale and the effects of synchrotron emission are non-negligible.

In this Letter, we use a semi-analytic model based on the solution of the Parker equation for the CR transport to calculate the CR proton and electron spectrum accelerated by typical SNRs, accounting for the effects of magnetic field amplification. We then use these spectra to estimate $q_{\rm e}$ and $q_{\rm p}$. This work represents the first calculation of the CR electron acceleration spectrum that self-consistently accounts for magnetic field amplification and thus the resulting synchrotron losses that take place within SNRs \cite[see, e.g.,][for examples of previous estimates]{ohira+12, berezhko+13}.
Our findings may have significant bearing on CR propagation models and the interpretation of observations such as the ``positron excess."

Let us now introduce the formalism that we use to model SNR evolution and CR acceleration.

\emph{Remnant Evolution---} SNRs are evolved using the formalism described in \cite{diesing+18}. 
More specifically, SNR evolution can be understood in terms of four stages: the \emph{ejecta-dominated stage}, in which the mass of the swept-up ambient medium is less than that of the SN ejecta, the \emph{Sedov stage}, in which the swept-up mass dominates the total mass and the SNR expands adiabatically, the \emph{pressure-driven snowplow}, in which the remnant cools due to forbidden atomic transitions but continues to expand because its internal pressure exceeds the ambient pressure, and, finally, the \emph{momentum-driven snowplow}, in which the internal pressure falls below the ambient pressure and expansion continues due to momentum conservation. 

While we model SNRs through the end of the pressure-driven snowplow, the majority of CRs are accelerated during the transition between the ejecta-dominated and Sedov stages. The DSA timescale for CRs of energy $E = \Emax$ is given by $\tau_{\rm DSA} \approx D/\vsh^2$ where $D$ is the diffusion coefficient and $\vsh$ is the shock speed. Assuming Bohm diffusion \citep{caprioli+14c}, $D(E) \propto \rl \sim E/B_2$ where $\rl$ is the Larmor radius and $B_2$ is the post-shock magnetic field. Thus, $\Emax \sim B_2\vsh^2 t$ and, during the ejecta dominated stage characterized by roughly constant velocity, $\Emax$ increases. During the Sedov stage, the shock slows down such that $\vsh \propto t^{-3/5}$, meaning that $\Emax$ decreases with time, i.e., $\Emax \propto B_2(t)t^{-1/5}$ \citep{cardillo+15,bell+13}. 
Our results are therefore most sensitive to the adiabatic SNR stages.

To model SNR evolution, we use the analytical approximation for the ejecta dominated stage presented in \cite{tm99}. Once the swept up mass exceeds the ejecta mass and the Sedov stage begins, we transition to the \emph{thin-shell approximation}, in which we assume that most of the mass resides in a thin layer that expands due to pressure in the hot cavity behind it \citep{bs95, om88, bp04}. 

All SNRs are assumed to eject $M_{\rm ej} = 1M_{\odot}$ (1 solar mass) with $E_{\rm SN} = 10^{51} \rm erg $ into a uniform ambient medium of density $\nism \in [10^{-2}, 10^{1}] \ \rm cm^{-3}$.

\emph{Proton Acceleration---}Instantaneous proton spectra are calculated using the Cosmic Ray Analytical Fast Tool (CRAFT) a semi-analytical formalism described in \cite{caprioli+10b,caprioli12} and references therein \citep[in particular,][]{ab05, ab06}. 
CRAFT self-consistently solves the diffusion-convection equation \citep[e.g.,][]{skilling75a} for the transport of non-thermal particles in a quasi-parallel, non-relativistic shock, including the dynamical backreaction of accelerated particles and of CR-generated magnetic turbulence. 
CRAFT is quick and versatile, but achieves the same degree of accuracy as Monte Carlo and numerical methods \citep{caprioli+10c}. 

Particles are injected into the acceleration mechanism following the prescription in \cite{bgv05}, namely that ions with momentum greater than $\xiinj$ ($ \sim $ a few) times the post-shock thermal momentum are promoted to CRs \citep[``thermal leakage," see][]{malkov98,kang+02}.
While kinetic simulations show that protons are injected via specular reflection and shock drift pre-acceleration rather than via thermal leakage \citep{caprioli+15}, such a prescription is calibrated with self-consistent kinetic simulations to ensure continuity between the thermal and non-thermal distributions \citep{caprioli+14a}.
$\xiinj$ can be mapped onto $\eta$, the fraction of particles crossing the shock injected into DSA, via
\begin{equation}
    \eta = \frac{4}{3\sqrt{\pi}}(R_{\rm sub}-1)\xiinj^3e^{-\xiinj^2},
\end{equation}
where $R_{\rm sub}$ is the subshock compression ratio (i.e., the ratio of the density immediately behind the shock to that immediately in front of it)  \cite{bgv05}. 
In this analysis, $\xiinj$ (and thus $\eta$) is left as a free parameter, which allows us to span a range of shocks where CRs are either test-particles or dynamically important.

Once the proton spectrum has been calculated at each timestep of SNR evolution, particle momenta are weighted by $1/L(t_0, t)$ and spectra are summed, with $L(t_0, t)\geq 1$ accounting for adiabatic losses \citep[see][for more details]{caprioli+10a,morlino+12}. 
Thus, we obtain a cumulative spectrum over the lifetime of the SNR. 
More specifically, $L(t_0, t)$ can be written in terms of the time-dependent decompression of a fluid element with initial density $\rho(t_0)\equiv\rho_0$. 
Since $\rho^\gamma(t) \propto \vsh^2(t)$,
\begin{equation}
    L(t_0, t) \equiv [\rho(t)/\rho_0]^{1/3} = [\vsh(t)/\vsh(t_0)]^{\frac{ 2}{3\gamma}},
\end{equation}
where $4/3\lesssim \gamma\lesssim 5/3$ is the adiabatic index of the plasma and CRs \citep[e.g.,][]{caprioli+10a, diesing+18}.

\emph{Magnetic Field Amplification---}The propagation of energetic particles ahead of the shock is expected to excite different flavors of \emph{streaming instability} \citep[]{bell78a,bell04,amato+09}, driving magnetic field amplification and enhancing CR diffusion \citep{caprioli+14b,caprioli+14c}. 
The result is magnetic field perturbations with magnitudes that can exceed that of the ordered background magnetic field. This magnetic field amplification has been inferred via the X-ray emission of many young SNRs, which exhibit narrow X-ray rims due to synchrotron losses by relativistic electrons \citep[e.g., ][]{parizot+06, bamba+05, morlino+10, ressler+14}. 

We model magnetic field amplification as in \cite{caprioli+09a,caprioli12}. 
Here, we assume that the pressure in Alfv\'{e}n waves saturates at $P_{\rm w} \simeq P_{\rm cr}/(2M_{\rm A})$, where $P_{\rm w}$ and $P_{\rm cr}$ are the pressures in Alfv\'{e}n waves and CRs normalized to the ram pressure $\rhoism \vsh^2$ and $M_{\rm A} \equiv \vsh/v_{\rm A} = \vsh \sqrt{4\pi\rhoism}/B$ is the Alfv\'{e}nic Mach number calculated in the amplified magnetic field. 
In the limit in which the fluid and Alfv\'{e}nic Mach numbers $\gg 1$, we obtain
\begin{equation}
    P_{\rm w}(x) = \frac{B(x)^2}{8\pi\rhoism\vsh^2} = \frac{1+u(x)}{4M_{\rm A}(x)u(x)}P_{\rm cr}(x),
\end{equation}
where $u(x)$ is the fluid velocity normalized to $\vsh$. Following the prescription described in \cite{morlino+12}, we find an expression for the magnetic field in front of the shock, 
\begin{equation}
    B_1 = \frac{\sqrt{4\pi\rho_1}u_1\vsh}{M_{\rm A, 1}} = \sqrt{\pi\rhoism}\vsh\frac{P_{\rm cr, 1}(2-P_{\rm cr, 1})}{1-P_{\rm cr, 1}}^{3/2},
\end{equation}
where the subscript 1 denotes quantities immediately in front of the shock. 
Behind the shock (denoted with subscript 2), the magnetic field strength is assumed to be $B_2 \simeq \sqrt{1+2R_{\rm sub^2}/3}B_1$, since magnetic field components perpendicular to the shock normal are compressed.
For $P_{\rm cr, 1}\approx 10\%$, the shock parameters described above give $B_2$ near a few hundred $\mu$G, in good agreement with X-ray observations of young SNRs \citep{volk+05,caprioli+08}.

\emph{Electron Spectrum---}Once the instantaneous proton spectrum, $f_{\rm p}(p)$, has been calculated, the instantaneous electron spectrum, $f_{\rm e}(p)$ is calculated as in \cite{morlino+09} using the analytical approximation provided by \cite{zirakashvili+07}:
\begin{equation}
   f_{\rm e}(p) =  K_{\rm ep}f_{\rm p}(p)\left[1+0.523(p/p_{\rm e, max})^{9/4}\right]^2e^{-p^2/p_{\rm e, max}^2},
\end{equation}
where $p_{\rm e, max}$ is the maximum electron momentum determined by equating the acceleration and synchrotron loss timescales. $K_{\rm ep}$ is the normalization of the electron spectrum relative to that of protons; its value ranges between  $10^{-2}$ and $10^{-4}$ \citep{volk+05,park+15,sarbadhicary+17} but has no bearing on the spectrum slope.

\begin{figure}[ht]
    \centering
    \includegraphics[trim=5px 10px 20px 20px, clip, width=0.48\textwidth]{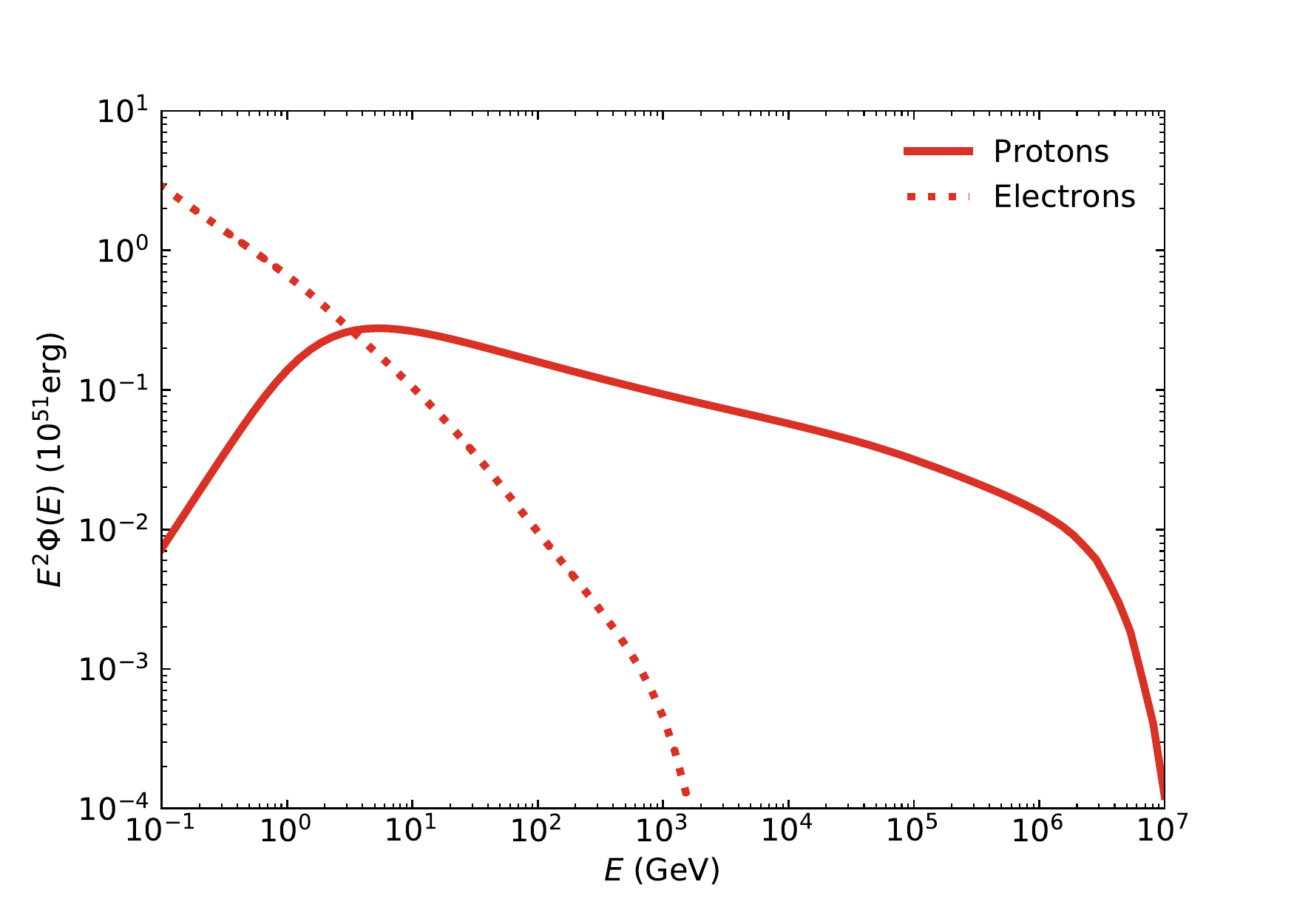}
    \caption{Cumulative proton and electron spectra near the end of the SNR lifetime. In this example, we take $\nism = 1.0 \rm \ cm^{-3}$ and $\xiinj = 3.3$ ($\eta \simeq 3\times10^{-3}$). Electron spectra are normalized by a factor of $1/K_{\rm ep}$ for display purposes.}
    \label{fig:SampleSpec}
\end{figure}

To determine the cumulative spectrum over the lifetime of an SNR, the electron energy $E$ is evolved by integrating
\begin{equation}
    \frac{dE}{dt} = -\frac{4}{3}\sigma_{\rm T} c\bigg(\frac{E}{m_{\rm e}c^2}\bigg)^2\frac{B_2^2}{8\pi}-\frac{E}{L}\frac{dL}{dt},
    \label{eq:eloss}
\end{equation}
where the first and second terms account for synchrotron and adiabatic losses respectively (inverse Compton losses are subdominant).
The weighted instantaneous spectra are then summed to determine a cumulative spectrum, an example of which shown in Figure \ref{fig:SampleSpec}. It is worth noting that electron escape upstream will have a negligible impact on this result, as escape is only important at the highest energies, where the acceleration time becomes comparable with the age of the system \citep{caprioli+10a}. When losses are important, the diffusion length of electrons will not allow them to escape.

{\it Results.---}
Having calculated the cumulative proton and electron spectra $\phi(E)$ at the end of the SNR lifetime, i.e., when the shock becomes subsonic and the remnant merges with the ISM, we estimate the power law slope as
\begin{equation}
    q\equiv -\left< \frac{d\log{\phi(E)}}{d\log{E}}\right>,
\end{equation}
 where $q$ is averaged between $10-10^4$ GeV for protons and between $10-100$ GeV for electrons.
 The energy ranges are chosen to ensure that particles are fully relativistic and to exclude high-energy cut-offs. The uncertainty in $q$ is estimated as the standard deviation over the range of calculation. 

\begin{figure}[ht]
\centering
\includegraphics[trim=70px 10px 540px 15px, clip, width=0.48\textwidth]{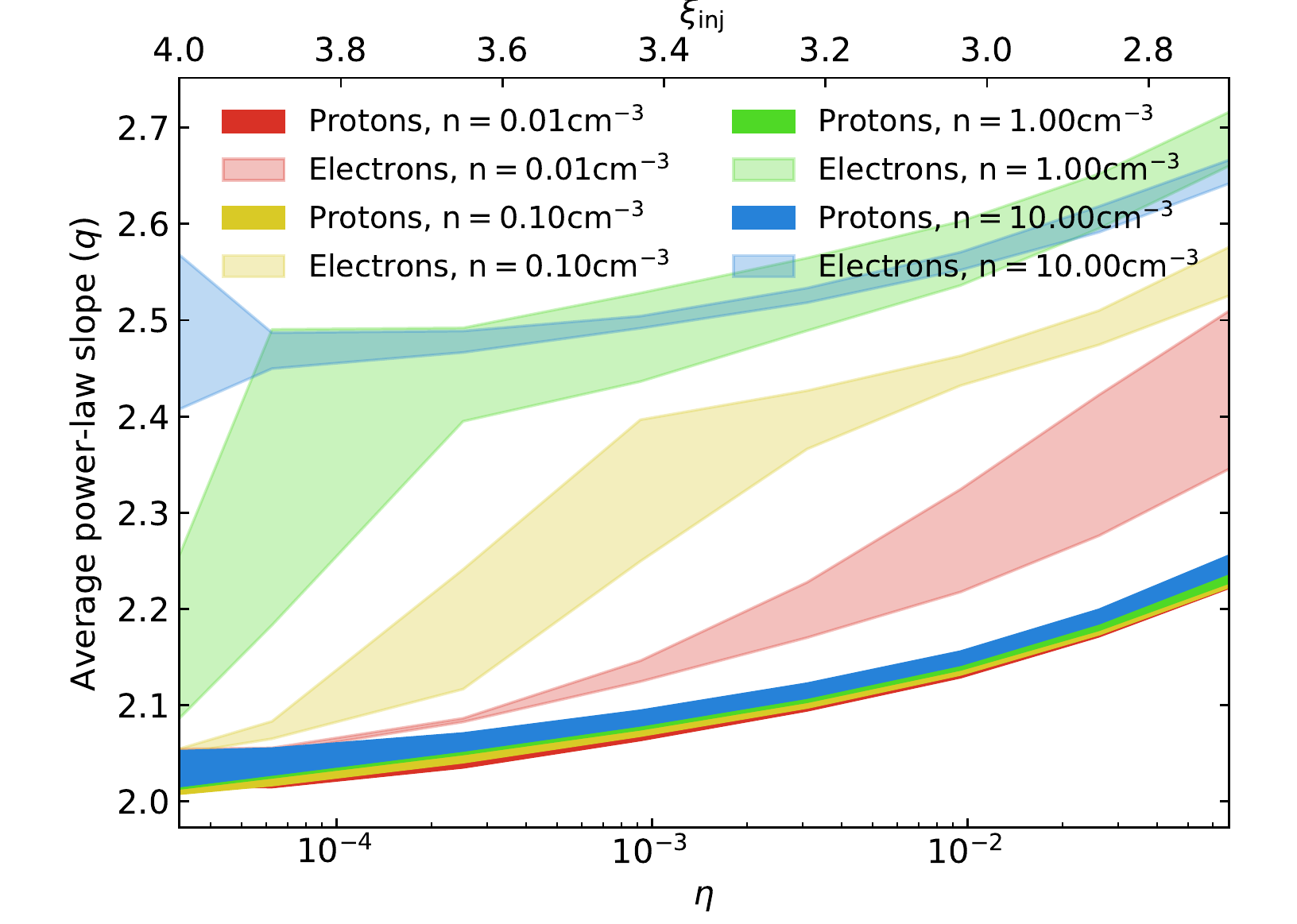}
\caption{Average slopes $q$ (where $f(E) \propto E^{-q}$) of modeled proton spectra (solid bands) and electron spectra (semi-transparent bands) at various ISM densities as a function of $\eta$ (see text for details). 
Electron spectra are consistently steeper than proton spectra, regardless of density or $\eta$. Note that the slopes of the proton spectra change little with density, meaning that some solid bands are hidden.}
\label{fig:FitResults}
\end{figure}

Figure \ref{fig:FitResults} shows the resulting average slopes of  electron and proton spectra  as a function of $\eta$ for multiple values of $\nism$. We find that the electron spectrum is consistently steeper than that of protons, regardless of acceleration efficiency or ISM density. 
For typical parameters ($n = 1 \ \rm{cm^{-3}}$ and $\eta \simeq 3\times10^{-3}$, which returns the canonical $P_{\rm cr}\simeq 10\%$ \citep{caprioli+14a}), $\Delta q\equiv q_{\rm{e}} - q_{\rm{p}} \simeq 0.4$. 

Figure \ref{fig:FitResults} also implies that $q_{\rm p}$, $q_{\rm e}$, and $\Delta q$ depend on $\eta$; when the number of particles injected into DSA increases, proton and electron spectra steepen. More specifically, in the low $\eta$ limit, we recover the ``test-particle scenario," in which the CR pressure is small and magnetic field amplification inefficient. The result is a proton slope consistent with the standard DSA prediction ($q_{\rm p} \simeq 2$). As $\eta$ increases, so too does the efficiency of magnetic field amplification and thus the velocity of magnetic perturbations responsible for scattering CRs. Since Alfv\'{e}n waves generated by CRs tend to travel against the fluid, this increase in magnetic field corresponds to a decrease in the effective compression ratio felt by CRs, resulting in a steepening of their spectrum \citep[see][]{zp08b,caprioli11,caprioli12}. 

An increase in $\eta$ also increases $\Delta q$, since larger $P_{\rm cr}$ and hence larger $P_{\rm B}$ lead to more severe synchrotron losses.
Increasing $\nism$ has a similar effect; the fraction of the bulk momentum flux converted to magnetic pressure is roughly constant at the few percent level such that $B_2\propto \rhoism^{1/2}\vsh$ (see Figure \ref{fig:FitResults2} for a clear illustration of this effect).

\begin{figure}[ht]
\centering
\includegraphics[trim=5px 10px 5px 10px, clip, width=0.48\textwidth]{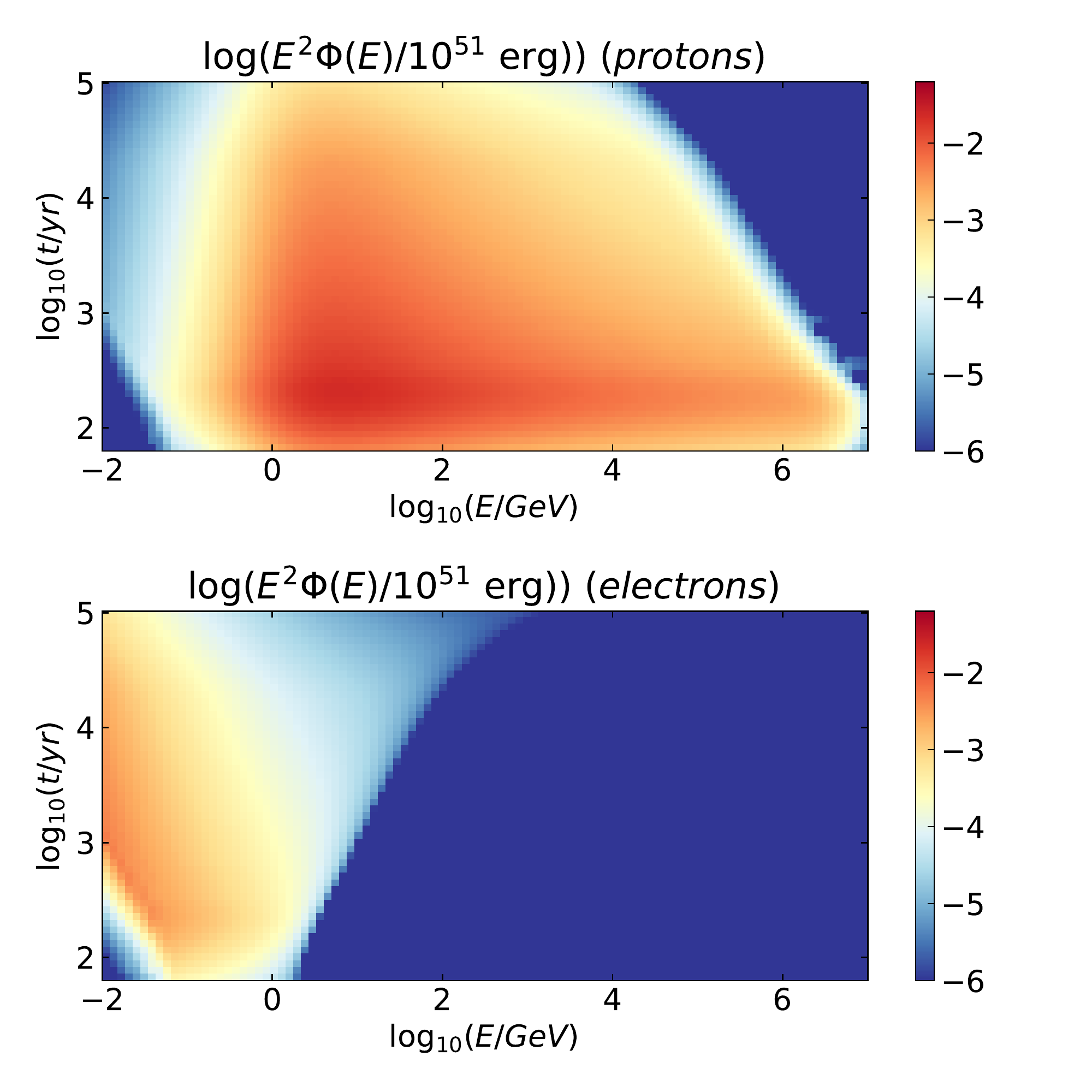}
\caption{Instantaneous proton (top) and electron (bottom) spectra, including energy losses, as a function of energy (x-axis) and time (y-axis). SNR parameters and electron normalization are the same as those in \ref{fig:SampleSpec}. The largest contribution to the proton spectrum occurs near the end of the ejecta-dominated stage ($t\sim 200$ yr).
The sharp steepening in high-energy electrons--which produces a distribution in time and energy that differs significantly from that of protons--reflects energy lost to synchrotron emission.
\label{fig:Contributions}}
\end{figure}

\begin{figure}[!t]
\centering
\includegraphics[trim=540px 10px 70px 15px, clip, width=0.48\textwidth]{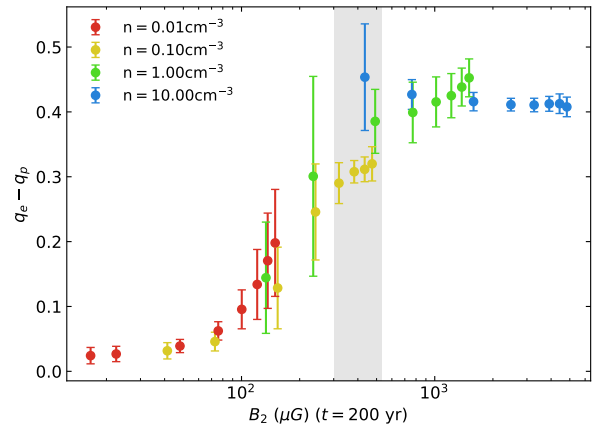}
\caption{Difference between electron and proton power law slopes ($\Delta q=q_{\rm e} - q_{\rm p}$) for various densities and values of $\eta$ as a function of magnetic field strength ($B_2$) at $t = 200$ yr (i.e., when the largest portion of CRs are accelerated; see Figure \ref{fig:Contributions}). While the onset of the Sedov stage (and thus the time at which the largest portion of CRs are accelerated) depends weakly on $n$, choosing $B_2$ at slightly earlier or later times has a negligible effect on the relationship between $B_2$ and $\Delta q$.
Due to synchrotron losses, stronger magnetic fields produce steeper electron spectra and thus larger $\Delta q$, with saturation at $\sim 0.4$. Gray fill indicates the post-shock magnetic field strength inferred for Tycho  \citep[]{morlino+12}.} 
\label{fig:FitResults2}
\end{figure}

Figure \ref{fig:Contributions} provides a more detailed picture of our calculated spectra and further illustrates the impact of synchrotron losses. 
The color scale indicates the magnitude of the instantaneous proton flux (top panel) and electron flux (bottom panel), weighted to account for energy losses, as a function of energy (x-axis) and time (y-axis). Note that the electron fluxes are multiplied by a normalization constant $1/K_{\rm ep}$ for display purposes. As suggested in the preceding section, the largest contribution to the proton spectrum occurs near the onset of the Sedov stage ($t\sim 200$ yr). 

To understand how Figure \ref{fig:Contributions} characterizes the synchrotron losses of CR electrons, recall that, in the case of protons, $\Emax$ is determined by equating the acceleration timescale, $\tau_{\rm DSA}$, with the lifetime of the remnant, $\tau_{\rm SNR}$, giving $\Emax \propto B_2(t)t^{-1/5}$ during the Sedov stage. This effect is apparent in Figure \ref{fig:Contributions} (top panel), which shows a clear cutoff at high energies that decreases with time. Electrons, on the other hand, experience synchrotron losses on a timescale $\tau_{\rm syn} < \tau_{\rm SNR}$, meaning that their $\Emax$ is set by $\tau_{\rm DSA} \simeq \tau_{\rm syn}$. 
Moreover, steepening (or rollover) of the electron spectrum will occur at an even lower energy, $E_{\rm roll}$, above which $\tau_{\rm syn} \lesssim \tau_{\rm SNR}$. 
Since $\tau_{\rm syn} \propto B_2^{-2}E^{-1}$, we find that $E_{\rm roll} \propto B_2^{-2}t^{-1}$. 
Assuming that $B_2\propto \vsh \propto t^{-3/5}$ during the Sedov stage, the result is that $E_{\rm roll} \propto t^{1/5}$. Again, this effect is apparent in Figure \ref{fig:Contributions} (bottom panel), which exhibits a sharp steepening in energy, the position of which increases with time. 

Since the difference between the CR electron and positron spectra arises from synchrotron losses, it is best understood in terms of the post-shock magnetic field, $B_2$ (see Figure \ref{fig:FitResults2}). 
Namely, as $B_2$ increases, so does $\Delta q$, since an increase in magnetic field strength leads to more severe synchrotron losses, and thus a steepening of the electron slope. 
Most notably, the post-shock magnetic field strengths inferred, e.g., for the Tycho SNR correspond to $0.3 \lesssim \Delta q \lesssim 0.4$ (gray band in Figure \ref{fig:FitResults2}), in perfect agreement with its multi-wavelength emission \citep{morlino+12}. 
In general, the typical magnetic fields that we estimate with CRAFT are consistent with those inferred from observations, implying that our conclusion that $\Delta q \gtrsim 0.3$ does not depend on modeling details of magnetic field amplification. 

{\it Discussion.---}
In summary, we used a semi-analytic model of non-linear diffusive shock acceleration to model the spectra of CR protons and electrons accelerated by SNRs. 
We find that electrons are injected into the Galaxy with spectra that are consistently steeper than those of protons, with a difference in slope of $\Delta q\simeq 0.1-0.4$. 
This steepening is the result of synchrotron losses in the large magnetic fields inferred in SNRs; therefore, it does not depend on the microphysics embedded in our model.

Our result may have significant implications for models of CR propagation in the Galaxy, which typically assume that protons and electrons are injected with the same spectrum.
In particular, it must be reckoned with the ``positron excess" reported by PAMELA \citep{pamela13} and AMS-02 \citep{ams14}. 
Comparisons between recent AMS-02 positron and electron data \citep{ams19a,ams19b}, and DAMPE and CALET electron+positron data \citep{dampe17,calet17} suggest that the positron fraction increases with energy as $\chi\propto E^{0.3}$ between 10 and $\sim 300$\,GeV.
In the standard propagation paradigm, $\chi(E)\propto E^{-(q_{\rm p}-q_{\rm e}+\delta)}\propto E^{\Delta q-\delta}$, hence $\Delta q\gtrsim \delta + 0.3$ could reproduce the rising in the positron fraction without introducing any source of primary positrons, be it astrophysical (e.g., pulsars) or exotic (dark matter).
Although measurements of CR lithium, beryllium, and boron suggest that $\delta \sim 0.2-0.4$ \citep[e.g.,][]{ams18}, the antiproton to proton ratio is instead consistent with $\bar{\delta} \lesssim 0.1$ \citep[][]{ams16a}. 
This discrepancy is likely due to the increase in the proton-proton cross section with energy, which partially compensates for diffusive steepening with a hardening of the antiproton spectrum, parameterized by $\epsilon$ \citep{donato+10,korsmeier+18}. 
Since positrons are produced in proton-proton interactions, too, taking $\bar{\delta} \sim \delta - \epsilon \lesssim 0.1$ also for the positrons implies that $\Delta q\simeq 0.4 \gtrsim \bar{\delta}+0.3$ can entirely account for the ``positron excess."
Note that, since the positron fraction is the ratio of lepton fluxes, radiative losses do not affect this conclusion.
Moreover, the expected secondary production due to propagation saturates the normalization of the positron spectrum \citep[e.g.,][]{blum+13}, while the positron to antiproton ratio is consistent with proton-proton branching ratios \citep[e.g.][]{lipari17,lipari19,blum+18}. Intriguingly, these findings suggest that positrons may be of secondary origin after all, and that the ``positron excess" may in fact be an \emph{electron deficit}.
This picture will be investigated more quantitatively in a forthcoming paper.

\begin{acknowledgments}
We thank E. Amato, F. Donato, and M. Korsmeier for their comments and discussions on secondary particle production.
This research was partially supported by NASA (grant NNX17AG30G and 80NSSC18K1726) and NSF (grant AST-1714658).
\end{acknowledgments}

\end{document}